\documentclass{article}

\usepackage{arxiv}

\usepackage[utf8]{inputenc} 
\usepackage[T1]{fontenc}    
\usepackage{hyperref}       
\usepackage{url}            
\usepackage{booktabs}       
\usepackage{amsfonts}       
\usepackage{nicefrac}       
\usepackage{microtype}      
\usepackage{lipsum}		
\usepackage{graphicx}
\usepackage{natbib}
\usepackage{doi}

\title{Cryptanalysis of authentication and key establishment protocol in Mobile Edge Computing Environment}

\author{ \href{https://orcid.org/0000-0002-1902-3910}{\includegraphics[scale=0.06]{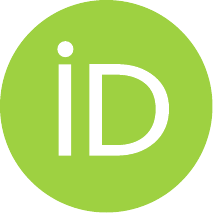}\hspace{1mm}Sundararaju Mugunthan} \\ 
	Department of Applied Mathematics and \\ Computational Sciences\\
	PSG College of Technology\\
	Coimbatore, Tamilnadu, India \\
	\texttt{mugunth05@gmail.com} \\
	\And
	\href{https://orcid.org/0000-0003-0691-0905}{\includegraphics[scale=0.06]{orcid.pdf}\hspace{1mm}Venkatasamy Sureshkumar} \\
	Department of Applied Mathematics and \\ Computational Sciences\\
	PSG College of Technology\\
	Coimbatore, Tamilnadu, India \\
	\texttt{sand.amcs@psgtech.ac.in} \\
}

\renewcommand{\shorttitle}{Cryptanalysis of authentication protocol in Mobile Edge Computing Environment}

\hypersetup{
pdftitle={Cryptanalysis of Wu et al.},
pdfsubject={cs.CR},
pdfauthor={Sundararaju Mugunthan, Venkatasamy Sureshkumar},
pdfkeywords={Cryptanalysis, Mobile Edge Computing, Authentication and Key Establishment protocol},
}

\usepackage{amssymb}
\usepackage{array,multirow}
\usepackage{amsthm}
\usepackage{pbox}
\usepackage{caption}
\usepackage{subcaption}
\usepackage{algorithmic}
\usepackage{algorithm}
\usepackage{amsmath}

\newcommand{\cmnt}[1]{}
\newcommand{\MUi}{MUser}
\newcommand{\PWi}{PW_m}
\newcommand{\MIDi}{MID_i}
\newcommand{\RC}{RC}
\newcommand{\TMIDi}{TMID_i}
\newcommand{\KRC}{R_S}
\newcommand{\Sj}{S_j}
\newcommand{\SIDj}{SID_j}
\newcommand{\PSIDj}{PSID_j}
\newcommand{\Ks}{K_s}
\newcommand{\rn}[1]{RN_#1}

\begin{document}
\maketitle

\begin{abstract}
	
Recently, in the area of Mobile Edge Computing (MEC) applications, Wu et al. proposed an authentication and key establishment scheme and claimed their protocol is secure. Nevertheless, cryptanalysis shows the scheme fails to provide robustness against key computation attack, mobile user impersonation attack and traceability attack. 
Vulnerabilities in their scheme lead to the exposure of mobile users' long term secret to mobile edge server provided both parties complete a successful session. 
This enables any malicious edge servers, who had communicated with the user earlier, to compute current session keys between the user and other legitimate servers. 
Also, since long term secret is exposed, such malicious servers can impersonate the user. We present a cryptanalysis of the scheme.
\end{abstract}


\section{Introduction}
The Mobile Edge Computing (MEC) is indispensable for providing computation and storage resources to time-critical and latency-sensitive applications. A robust security protocol with necessary security features is essential to protect communications in such applications.
The authors of the protocol claimed their scheme is resistant to Man-in-the-middle attack, replay attack, offline password guessing attack, user impersonation attack and insider attack. Unfortunately, their scheme is vulnerable to many security attacks. In the following section, We provide a cryptanalysis of their protocol.

\section{Cryptanalyis of Wu et al.}
\label{cryptanal}

The authors \cmnt{Wu et al.} \cite{wu2022provably} presented a security protocol to enable mutual authentication and key establishment among mobile users, registration center and mobile edge servers. The authors claimed their scheme resists security attacks. However, their scheme is found vulnerable to many security attacks. In this section, the overview of their protocol is presented and subsequently the vulnerabilities are shown. 

\subsection{Overview of Wu et al.}
The scheme consists of two phases namely registration phase, authentication and key establishment phase. In registration phase, mobile users and edge servers register with the registration center using a secure channel. The AKE phase uses insecure public channel for communication.

\subsubsection{Mobile User registration phase}

\noindent \textbf{Step R1:} The Mobile user $\MUi$ chooses an identity $\MIDi$ and sends it to the $\RC.$

\noindent
\textbf{Step R2:} The $RC$ generates random numbers $r_i$ and $x_i$, computes $\TMIDi = h(\MIDi \| r_i)$, $B_i = \TMIDi \oplus h(\KRC \|  x_i)$ and the $\RC$ stores the values $\{B_i, x_i\}$ in its database. Also, the $\RC$ generates a smart card with following values $ SC = \{\TMIDi, B_i, h(\cdot) \}$ and sends it to the mobile user using a secure channel. 

\noindent \textbf{Step R3:} $\MUi$ inputs password $\PWi$ and imprints the biometrics $Bio_m$, $\MUi$ generates a random number $n_i$, computes $Gen(Bio_m) = (\sigma_i, \tau_i)$, $C_i = n_i \oplus h(\MIDi \| \PWi \| \sigma_i)$, $Auth_i = h(\MIDi \| \PWi \| \sigma_i \| n)$, and $\TMIDi^* =  \TMIDi \allowbreak \oplus h(n \| \PWi \| \sigma_i)$, and then replaces $\{ \TMIDi \}$ with $\{ \TMIDi^* \}$ and stores $\{C_i, \allowbreak Auth_i, \allowbreak Gen(\cdot), Rep(\cdot), \tau_i \}$ in a smart card $SC_i.$

\subsubsection{Server Registration Phase}
Registration of the mobile edge server with the Registration Center realised using following steps. 

\noindent \textbf{Step M1:} The server $\Sj$ generates an identity $\SIDj$ and forwards it to the $\RC$ through a secure channel. 
    
\noindent \textbf{Step M2:} The $\RC$ generates a random number $r_j$ and $x_j$ and it computes the following values $\PSIDj = h(\SIDj \| r_j)$, $Q_j = h(\SIDj \| x_j)$, and $F_j = r_j \oplus Q_j.$ The master secret of the edge server is computed as $\Ks = h(\SIDj \| \KRC)$.  The values $\{F_j, \PSIDj, \allowbreak x_j\}$ are stored in $RC$'s memory and shares the values  $\Ks$ and $r_j$ with $\Sj.$

\noindent \textbf{Step M3:} On receiving the values $\{\Ks, r_j \}$, the server $\Sj$ stores the data in its database. 

\subsubsection{Login and authentication phase}
In this phase, a mobile user $\MUi$ establishes a secure session with a edge server using the help of the $\RC.$ The steps involved are as follows. 

\noindent \textbf{Step L1:} Mobile user $\MUi$ inserts his/her smart card $SC_i$ and inputs the values $\MIDi, \PWi$, presents his/her $Bio_m', \MUi$, computes $Rep(Bio_m', \tau_i) = \sigma_i'$, $n_i' = C_i \oplus h(\MIDi \| \PWi \| \sigma'_i)$. Also, $\MUi$ verifies whether $Auth_i \overset{?}{=} h(\MIDi \| \PWi \| \sigma'_i \| n'_i)$ holds. If the verification is successful, $\MUi$ generates a random number $\rn1$ and timestamp $TS_U$, computes $ \TMIDi = \TMIDi^* \oplus h( n_i \| \SIDj \| \TMIDi)$, $\rn1' = \rn1 \oplus h(\SIDj \| \TMIDi)$, $D_1 = \SIDj \oplus h(\TMIDi \| \allowbreak TS_U)$, $D_2 = h(\SIDj \| \TMIDi \| \rn1 \| TS_U)$, and then $\MUi$ sends the message $M_1 =\{ \rn1', B_i, D_1, D_2, TS_U \}$ to $\RC.$

\noindent \textbf{Step L2:} When the $\RC$ receives $M_1,$ it checks the time delay by computing $| TS_U - T_C | \leq \Delta T$, and if its is in permissible limit, the $\RC$ computes $\TMIDi = B \oplus h(\KRC \| x_i)$, $\SIDj = D_1 \oplus h(\TMIDi \| TS_U)$, $\rn1 = \rn1' \oplus h(\SIDj \| \TMIDi \| \rn1 \| TS_U)$, and verifies $D_2' \overset{?}{=} D_2;$ if the verification fails, the $\RC$ closes the session. Else, the $\RC$ retrieves $\SIDj$'s $\PSIDj$ and generates a random number $\rn2$ and timestamp $TS_{RC}.$ Then, the $\RC$ computes $D_3 = h(\PSIDj \| \rn2) \oplus h(h(\SIDj \| \KRC))$, $Q_j = h(\SIDj \| x_j$, $r_j = F_j \oplus Q_j$, $\TMIDi' = h(\PSIDj \| r_j) \oplus \TMIDi,$ $D_4 = h(h(\PSIDj \| \rn2) \| \SIDj \| TS_{RC})$. Then, the $\RC$ sends $M_2 = \{\rn1', TS_{RC}, \TMIDi', D_3, D_4 \}$ to $\Sj.$

\noindent \textbf{Step L3:} On receiving the message  $M_2$, the edge server $\Sj$, checks the validity of the message by checking $|TS_{RC} - T_C | \leq \Delta T$, and when it holds, $\Sj$ computes $h(\PSIDj \| \rn2) = D_3 \oplus h(\Ks)$ and $D_4' = h(h(\PSIDj \| \rn2) \| \SIDj \| TS_{RC})$, and the $\RC$ checks $D_4' \overset{?}{=} D_4.$ If correct, $\Sj$ generates a random number $\rn3$ and timestamp $TS_{MS}$ and computes $\PSIDj = h(\SIDj \| r_j)$, $\TMIDi = \TMIDi' \oplus h(\PSIDj \| r_j),$ $\rn1 = \rn1' \oplus h(\SIDj \| \TMIDi)$,  $SK = h(h(\PSIDj \| \allowbreak \rn2) \| \rn1 \| \allowbreak \rn3 \| \TMIDi)$, $\rn3' = \rn3 \oplus  h(\allowbreak \rn1  \| \SIDj)$, $D_5 = h(\rn1 \| \rn3 \| \allowbreak TS_{MS})$, and $D_6 = h(SK \|  h(\allowbreak \PSIDj \| \rn2) \| \rn1).$ Then, $\Sj$ sends the message $M_3 = \{ TS_{MS}, \allowbreak \rn3', D_5, D_6 \}$ to $\Sj.$

\noindent \textbf{Step L4:} When message $M_3$ is received, the $\RC$ calculates the time difference as $|TS_{MS} - T_C| \leq \Delta T.$ If it is within the permissible limit, the $\RC$ selects a timestamp $T_4,$ computes $\rn3 = \rn3' \oplus h(\rn1 \| \SIDj)$ and $D_5' = h(\rn1 \| \rn3 \| TS_{MS})$, and verifies $D_5' \overset{?}{=} D_5$. If correct, the $\RC$ computes $D_7 = h(\PSIDj \| \rn2) \oplus h(\rn1 \| \TMIDi)$ and $D_8 = h(h(\PSIDj \| \rn2) \| \rn3 \| T_4)$, and sends the message $M_4 = \{ T_4 \| \rn3', D_6, D_7, D_8 \}$ to the user $\MUi$.

\noindent \textbf{Step L5:} The user $\MUi$, on receiving the message $M_4$, verifies the validity of the message by computing $ |T_4 - T_C | \leq \Delta T $, and if valid, $\MUi$ computes $\rn3 = \rn3' \oplus h(\rn1 \| \SIDj)$, $h(\PSIDj \| \rn2) = D_7 \oplus h(\rn1 \| \TMIDi)$ and $D_8' = h(h(\PSIDj \| \allowbreak \rn2) \|  \rn3 \| T_4),$ and verifies $D_8' \overset{?}{=} D_8;$ if correct, $\MUi$ constructs the session key as $SK = h(h(\PSIDj \| \rn2) \| \rn1 \| \rn3 \| \allowbreak \TMIDi)$, and $D_6' = h(SK \| h(\PSIDj \| \rn2)$ and verifies $D_6' \overset{?}{=} D_6$. If true, $\MUi$ saves $SK$ for future communication. 

\subsection{Pitfalls in Wu et al. scheme}
Although the authors claimed their protocol is secure, the scheme by \cmnt{Wu et al.} \cite{wu2022provably} is unable to resist key computation attack, user impersonation attack, traceability attack and honest but curious registration centers. Execution of these attacks are presented below.

\subsubsection{Key computation attack}
The attack is executed under the following circumstance. 
Suppose that a user successfully completes authentication and key establishment with the edge server $SID_j$, namely in session $n.$ After disconnecting the communication with the server $SID_j,$ the mobile user initiates authentication and key establishment with another edge server $SID_j^*$ in the session $(n+k)$.
In this communication, $SID_j$ can act as an attacker and compute the common session key that is about to be established among the mobile user, edge server $SID_j^*$ and the Registration center $RC.$
The main components involved in the construction of session key are $h(PSID_j \|\rn2), \rn1, \rn3$ and $TMID_i$. If an attacker able to compute these components from a random session, then the attacker can compute the associated key of the session. \\

\noindent Following steps are executed to perform the attack. \\

\noindent \textbf{Step KC1:} 
During the AKE between user and $SID_j,$ the edge server $SID_j$ comes to know $TIMD_i$ in step 3 of the \cmnt{Wu et al.} \cite{wu2022provably} protocol as $TMID_i = TMID_i' \oplus h(PSID_j \| r_j)$ where $PSID_j$ is the masked identity of $SID_j$ and $r_j$ is its secret. 
$SID_j$ retains this $TMID_i$ to execute session key computation attack in the authentication and key establishment process among the mobile user, $SID_j$ and $RC$ as follows.

\noindent \textbf{Step KC2:} 
In the session $(n+1),$ during the exchange of message $M_1^* = \{\rn1'^*, B_i, D_1^*, D_2^*, TS_U^* \}$, $SID_i$  captures $M_1^*$ and computes $\rn1^*$ as follows. 

\noindent
For this purpose, $SID_j$ extracts primarily $SID_j^*$ from $D_1^*$ as $SID_j^* = D_1^* \oplus h(TMID_i \| TS_U^*)$. Using the values $SID_j^*$ and $TMID_i$ in $\rn1'^*$, the $\rn1^*$ is extracted as $\rn1^* = \rn1'^* \oplus h(SID_j^* \| \allowbreak TMID_i).$

\noindent \textbf{Step KC3:} 
Then, in the step 4 of the session $(n+1),$ the attacker extracts the value $h(PSID_j^* \| \rn2^*)$ from $D_7^*$ of the message $M_4^* =\{T_4^*, \rn3'^*, D_6^*, D_7^*, D_8^* \}$ as $h(PSID_j^* \| \rn2^*) = D_7^* \oplus h(\rn1^* \| TMID_i).$ 

\noindent \textbf{Step KC4:} 
From the same message $M_4^*,$ the attacker $SID_j$ computes $\rn3^*$ from $\rn3'^*$ as $\rn3^* = \rn3'^* \oplus h(\rn1^* \| SID_j^*)$

\noindent \textbf{Step KC5:} 
Now the attacker has all the primitive components required for the construction of the session key $SK^*$ of the session $(n+1)$ and thus the session key $SK^*$ is computed as $SK^* = h(h(PSID_j^* \| \rn2^*) \| \rn1^* \| \rn3^* \| TIMD_i).$
Thus the scheme is vulnerable to session key computation attack. 

\subsubsection{Mobile User impersonation attack}
Under the same circumstances mentioned in the above attack, $SID_j$ can impersonate $TIMD_i$ as follows.

Suppose that a user successfully completes authentication and key establishment with the edge server $SID_j$, namely in session $n.$ After disconnecting the communication with the server $SID_j,$ the mobile user initiates authentication and key establishment with another edge server $SID_j^*$ in the session$~(n+k)$. \\

\noindent However, with the help of the credentials shared in the session $n$, edge server can impersonate the user. 

\noindent
\textbf{Step UI1:} 
During the authentication and key establishment between user and $SID_j,$ the edge server $SID_j$ comes to know $TIMD_i$ in step 3 of the Wu et al. protocol as $TMID_i = TMID_i' \oplus h(PSID_j \| r_j)$ where $PSID_j$ is the masked identity of $SID_j$ and $r_j$ is its secret. $SID_j$ retains this $TMID_i$ to impersonate the mobile user $\MUi$ by producing a new access request message $M_1$ as follows. 

\noindent
\textbf{Step UI2:} 
$\MUi$ selects a random $\rn1 \oplus h(\SIDj \| \TMIDi)$, $D_1 = \SIDj \oplus h(\TMIDi \| TS_U)$, $D_2 = h(\SIDj \| \TMIDi \| \rn1 \| \allowbreak TS_U)$, and then $\MUi$ sends the message $M_1 =\{ \rn1', B_i, D_1, \allowbreak D_2, TS_U \}$ to $\RC.$

\noindent
\textbf{Step UI3:}
On receiving the message $M_1,$ the $\RC$ checks the time delay by computing $| TS_U - T_C | \leq \Delta T$, and if its is in permissible limit, the $\RC$ computes $\TMIDi = B \oplus h(\KRC \| x_i)$, $\SIDj = D_1 \oplus h(\TMIDi \| TS_U)$, $\rn1 = \rn1' \oplus h(\SIDj \| \TMIDi  \| \allowbreak \rn1 \| TS_U)$, and verifies $D_2' \overset{?}{=} D_2.$ Since it is true, the $\RC$ accepts the message, authenticates $\SIDj$ as $\MUi$ and proceeds to key establishment phase. Thus, $\SIDj$ succeeds in impersonating the mobile user. 

\subsubsection{Traceability attack}
The mobile user $\MUi$ produces the message \(M_1 = \{\rn1', \allowbreak B_i, D_1, D_2, TS_U \} \). Here, the components \(\rn1', D_1,  D_2, TS_U\) are different for distinct sessions. However, the component $B_i$ remains constant in all session. Although it is not an identity information and prevails user anonymity, it enables a passive attacker to trace the actions of \(\MUi\). Thus, this protocol is vulnerable to traceability attack. 

\subsubsection{Honest but curious registration center}
The authors have mentioned in the article that the requirement of the scheme as key establishment between the mobile user and edge server. However, the designed protocol by \cmnt{Wu et al.} \cite{wu2022provably} establishes the session key among the three entities - the mobile user, edge server and registration center. Also, as a passive attacker, it is possible for the $RC$ to obtain the session key $SK$. It is noteworthy, that the information on further communication between the mobile user and edge server is revealed to $RC$ unnecessarily. 

\section{Conclusion}
We present a cryptanalysis of the scheme by Wu et al. to show the protocol is vulnerable to impersonation attack, key computation attack and untraceability attack. 
As a future work, this protocol can be enhanced such that the improved scheme is free from these vulnerabilities. 

\bibliographystyle{unsrtnat}

\bibliography{references}  






\end{document}